# Experimental Estimation of Temporal and Spatial Resolution of Coefficient of Heat Transfer in a Channel Using Inverse Heat Transfer Method

Majid Karami [1], Somayeh Davoodabadi Farahani [2], Farshad Kowsary [1*], Amir Mosavi [3,4*]

[1] School of Mechanical Engineering, University of Tehran, Tehran, Iran; fkowsari@ut.ac.ir & majid.karami85@gmail.com

[2] School of Mechanical Engineering, Arak University of Technology, Arak, Iran; sdfarahani@arakut.ac.ir

[3] School of the Built Environment, Oxford Brookes University, Oxford OX3 0BP, UK; a.mosavi@brookes.ac.uk

[4] Kando Kalman Faculty of Electrical Engineering, Obuda University, 1034 Budapest, Hungary; amir.mosavi@kvk.uni-obuda.hu

*Abstract*

In this research, a novel method to investigation the transient heat transfer coefficient in a channel is suggested experimentally, in which the water flow, itself, is considered both just liquid phase and liquid-vapor phase. The experiments were designed to predict the temporal and spatial resolution of Nusselt number. The inverse technique method is non-intrusive, in which time history of temperature is measured, using some thermocouples within the wall to provide input data for the inverse algorithm. The conjugate gradient method is used mostly as an inverse method. The temporal and spatial changes of heat flux, Nusselt number, vapor quality, convection number, and boiling number have all been estimated, showing that the estimated local Nusselt numbers of flow for without and with phase change are close to those predicted from the correlations of Churchill and Ozoe (1973) and Kandlikar (1990), respectively. This study suggests that the extended inverse technique can be successfully utilized to calculate the local time-dependent heat transfer coefficient of boiling flow.

**Introduction**

Boiling flow of a tube takes place in a variety of equipment such as power plants (steam, solar, and cooling), chemical industries, refrigeration, and air conditioning systems, rendering accurate knowledge of this phenomenon important. In a two-phase flow, the relative distribution of the liquid phase and gas in the tube flow is required to describe the flow. The pattern of boiling flow in the horizontal and vertical pipes is similar. Their difference lies only in the effect of gravity on the flow, which tends to remain liquid and gas in the down and above of the channel, respectively. Many works have been conducted on different criteria of boiling [1-3].

Tibiriça and Ribatski [4] performed experimental tests to study boiling heat transfer of R245a in a tube. Bodhal [5] studied bubbly boiling flow regime in the channel and determined its wavy character. In the rectangular narrow channel, Kim and Mudawar [6] found out that in addition to the convection, nucleate boiling has a considerable result in the boiling heat transfer. Boiling heat transfer of hydrocarbon fluids in a vertical tube is experimentally studied by Wadekar [7]. He observed pressure drop had been affected by subcooled boiling in the region where almost the relative distribution of gas to liquid is close to zero. Aria et al. [8] investigated experimentally heat transfer of R-134a boiling inside a spiral coiled pipe and realized that when the tube changes from straight to spiral, as well as the pressure drop, heat transfer is also increased. In addition, there have been extensive research to understand the effect from adding nanoparticles: $Al_2O_3$ water [9], CuO water [10], and Ag water [11] to fluids in boiling flow. Boudouh et al. [12] studied the boiling flow with nanofluid in mini channel, finding out that Nusselt number and vapor quality increased with nanofluid, in comparison to the base fluid. Chehade et al. [13] investigated boiling thermal efficiency of silver nanoparticle in water flow in micro channel. They used two series of thermocouples in 0.5mm and 8.5mm from the internal surface channel and calculated local heat flux by means of measured temperature and Fourier Law. They found out that boiling heat transfer in the channel entrance was more enhanced. Huang et al. [14] studied transient thermal analysis of electronic tools under a variable heat flux in a multi-micro-channel evaporator. They considered two and three-dimensional thermal models. Nonetheless, very few works have been performed, dedicated to the local time-dependent changes of the Nusselt number and the quality of vapor. Therefore, it is important to attempt understanding thermal phenomena such as boiling flow. Keeping these in mind, an inverse method has been developed to estimate time history of local heat flux input into the fluids in a rectangular channel. The objective in the inverse problem is to predict the unknown using the temperatures measured within the body. These unknowns [15] can be temperature, thermal flux, thermal properties, heat source or part of the body's geometry and inverse design [16].The measured temperature data are with noise, while inverse heat conduction problems are sensitive to noise. Thus, these problems are ill-posed.

Regularization techniques are utilized to alleviate this problem [17-19]. The inverse problem of predicting heat flux is a linear problem. One of the regularization methods [20] is the conjugate gradient method. Farahani and Kowsary [21] estimated local Nusselt number in mini-channel numerically, utilizing the inverse technique. Giedt [22] investigated variations in the Nusselt number of the airflow perpendicular to a cylinder using an inverse technique. Abia and Yamazkai [23] conducted an experiment to investigate changes in heat transfer using an inverse method in a series of cylinders exposed to airflow. Taler [24] predicted the values of the Nusselt number on the pipe circumference by using inverse method.

The heat transfer in the twisted pipes is investigated using the inverse technique, in the fully developed laminar flow by Borzoi et al. [25] and in the transitional regime by Cattani et al. [26]. Rouizi et al. [27] predicted the bulk temperature in a mini channel using two inverse techniques.

Closer examination of the boiling phenomenon needs to find a precise distribution of heat flux on the channel's surface. To achieve this goal, the inverse technique has been proposed in this study. This method, a non-intrusive one, which is based on conjugate gradient technique, has been utilized to predict heat flux into the fluids on heat exchange surface, using K-type thermocouples, inserted in specific locations within the channel, to measure time history of the temperatures, therefore, it provides inputs for the inverse procedure. For solving the heat equation, the finite element technique in ANSYS commercial software has been utilized with inverse algorithm code, written in ANSYS parametric design language (APDL). The flow with and without phase change is considered in the channel for investigation. Transient two-dimensional thermal model is utilized to study the heat flux, surface temperature, Nusselt number, vapor quality, convection number, and boiling number.

**Test Set up**

A plan of the test set up in this research is illustrated in figure1. The fluid in the tank is pumped and the impurities are separated from the fluid by a filter and a regulating valve is employed to set the mass flux. The uncertainty in the measured flow rate is 5%. In the flow with phase change, Outflow from the test section is in the two - phase state and before returning to the tank, the heat exchanger cools the working fluid and become liquid.

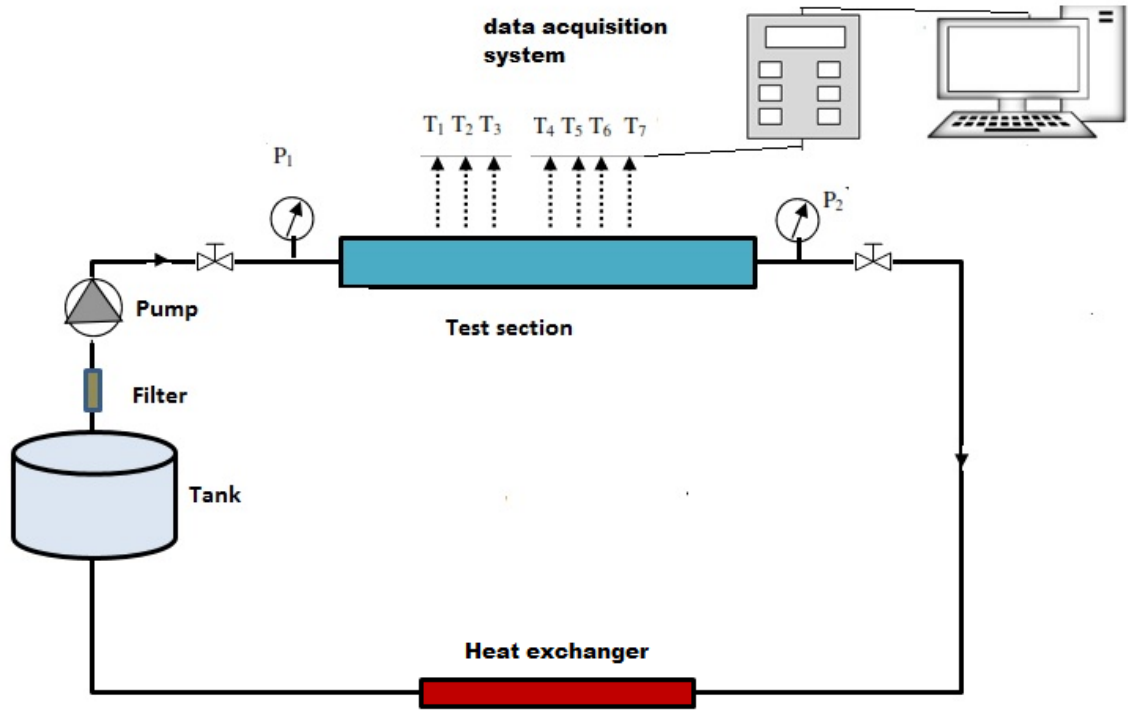

**Figure 1** A schematic diagram of the experiment set up

The test section (Figure 2) is a stainless steel duct, its length is 1.2m. Its cross-section is a rectangle (40×25 mm$^2$). Each wall, 10 mm thick, is heated by a rectangular heating panel to provide uniform heating. To reduce heat dissipation, three layers of thermal insulation have been taken into consideration. The first layer is asbestos with a thickness of 5mm ($k_{asbestos}$= 0.15 W/ (mK)) which is directly positioned on the thermal heater; the second layer is fiberglass with a thickness of 30mm (with $k_{fiberglass}$= 0.04 W/ (mK)) which is applied around the duct; and finally the third layer is nylon tape (2 mm thick and 50 mm wide with thermal conductivity of 0.25 W/ (mK)) to hold all the other layers together. These layers of insulation ensure that heat losses are kept below 5%. The total power of the input heating panels is 3000 W. A power supply has been used to control the input heating panel. Power supply accuracy is about 1%. The inner wall temperatures are measured with seven K-type thermocouples, located along the duct at axial distances of 50, 150, 350, 600, 950, 1050, and 1150 mm from the duct inlet. They are placed inside the holes and drilled into the duct, 0.5 mm from the upper surface.

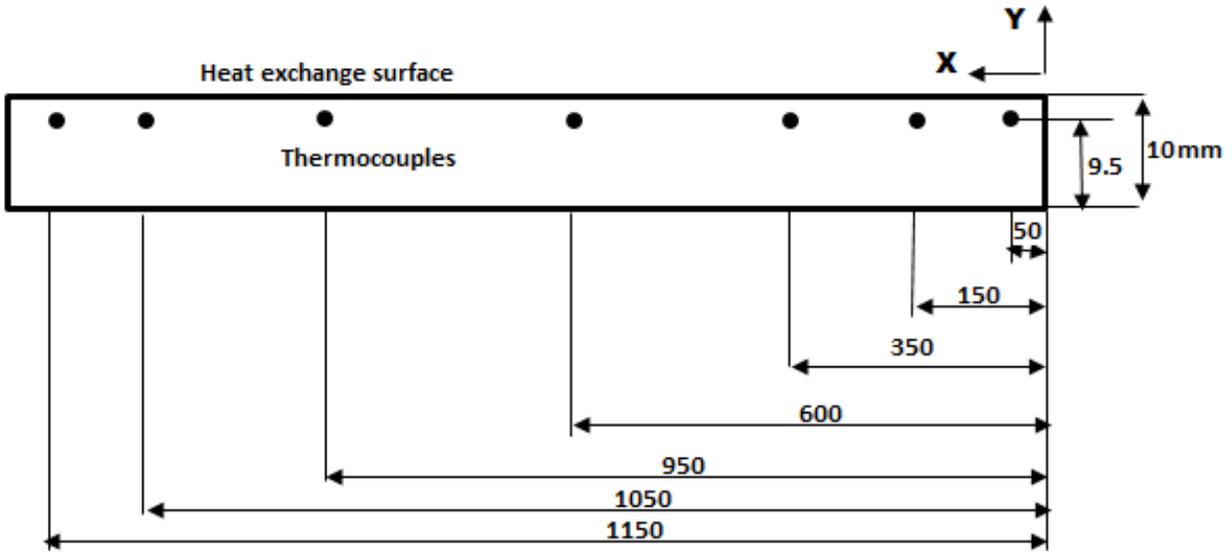

**Figure 2** Thermocouple locations in the channel (The dimension is mm)

Table 1 shows the uncertainty associated with measurement of devices, such as flow meters, thermocouples, heater power, and machining. Applying the proposed method in [28], it is estimated that the general uncertainty for measuring Nusselt numbers is within the range of 5.8 ± 0.8%.

**Table.1 The error source**

| Error source | Bias |
|---|---|
| Thermocouple | 0.1°C |
| Flow meter | 5% |
| Heater power | 1% |
| Machining process | 0.02mm |

**Problem Description**

The conjugate gradient technique has been employed to predict heat fluxes into the fluid in a rectangular channel. The channel is made of AISI 313. The heat flux is equal to the heater power. Temporal and spatial variations of heat flux and surface temperature at y=E is unknown. It has been assumed that the loss of heat is negligible with the governing equation of the plate[20], written as:

$$\begin{aligned}
&T_{xx} + T_{yy} = \alpha^{-1} T_t \\
&T_x \big|_{x=0,L} = 0 \\
&T_y \big|_{y=0} = -q_w k^{-1} \\
&T_y \big|_{y=E} = -q(x,t) k^{-1} \\
&T(x,y,0) = T_i
\end{aligned} \quad (1)$$

Where $T_i$, L, E, and $q_w$ are the initial temperature, length and thickness of plate, and heater power, respectively. $q(x,t)$ is unknown; thus instead of the heat flux, the measured temperatures by the thermocouples are available from the experimental test. The temperature of the channel wall is affected by the changes in the Nusselt number. We used this matter and solved the problem to predict indirectly the convective heat transfer coefficient (h). First, the $q(x,t)$ is estimated and then $h$ is estimated by using Newton's cooling law and estimated $q(x,t)$. Thus, we modeled the heat conduction equation (equation (1)) in the channel wall and convection term has appeared in the boundary condition. This term is unknown. The measured data contains information about $h$, which can be achieved by decoding method. The conjugate gradient method as the inverse technique has been used to retrieve this information. The temperatures are measured inside the plate using the thermocouples. These measurements are as input in the inverse technique. The most important point of this technique is that no information is required from the fluid flow field.

**Table 2** Thermal property of work material

| Thermal property | value |
|---|---|
| Specific heat, $C_p$ (J/(kgK)) | 460 |
| Thermal conductivity, $k(w/(mK))$ | 28.6 |
| Density, $\rho(kg/m^3)$ | 7760 |

**Table 3** Thermocouple position coordinates

| Thermocouple no. | X (mm) | Y (mm) |
|---|---|---|
| T1 | 50 | -0.5 |
| T2 | 150 | -0.5 |
| T3 | 350 | -0.5 |
| T4 | 600 | -0.5 |
| T5 | 950 | -0.5 |
| T6 | 1050 | -0.5 |
| T7 | 1150 | -0.5 |

In order that heat equation solving, the Finite-Element Method has been used in ANSYS software. In the direct thermal problems, all boundary conditions and material properties (Table 2) are known. Table 3 shows the location of thermocouples within the plate. The inverse method, conjugated gradient method (CGM), estimates the $q(x,t)$, using the temperatures are recorded from specified positions in the plate. The details of the CGM algorithm has been presented in pervious author's work [20], thus they will not be brought here. The inverse method employs a parameter, called the sensitivity coefficient, in fact indicating the temperature sensitivity of the unknown parameter. The sensitivity coefficient [16] is calculated as:

$$X(x_m, y_m, t) = \frac{\partial T(x_m, y_m, t)}{\partial q} \quad m = 1,2,..,Ns \tag{2}$$

where $x_m$ and $y_m$ are the sensor location within the channel wall, and P is the number of sensors. It must be noticed that the number of unknowns should be smaller or equal to the number of thermocouples in IHCP [15]. No functions have been employed to predict the heat flux components. One of the principles of inverse heat transfer is that we have no knowledge about the unknowns. The flow chart of which is illustrated in Figure 3, showing the iterative procedure for an inverse method. In this study utilizes seven thermocouples. At least one thermocouple is required to estimate each heat flux component in IHCP [20]. In order to describe heat flux on the heat exchange surface, this study estimates seven parameters, $q_1$, $q_2$…, and $q_7$ which vary with time. Error analysis is done to determine the accuracy of the proposed algorithm. As such, this analysis is performed, considering AISI304 as plate material and $q_w = cte$. Moreover, a known heat flux at the heat exchange surface (i.e. $q_{exact}$) and initial work piece temperature of 25°C.

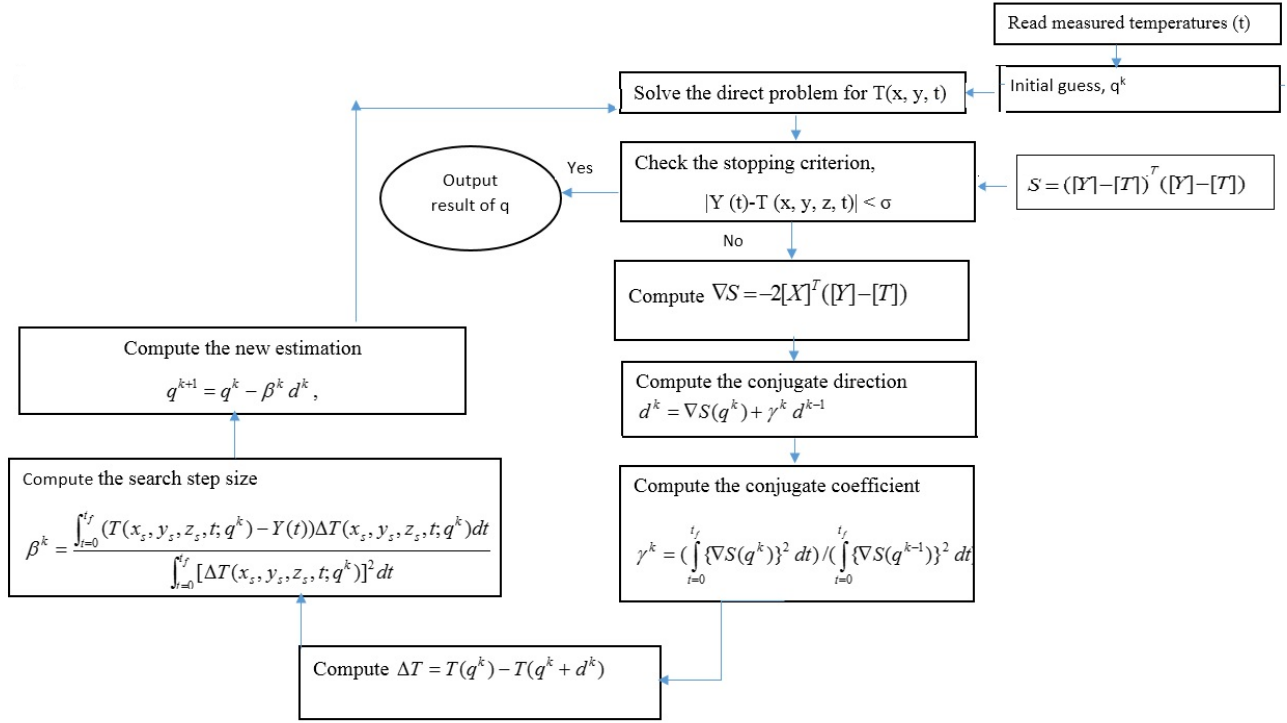

**Figure 3** Inverse algorithm

Temperatures, obtained as a result of a known imposed $q_{exact}$, are perturbed by a noise with (σ=0.1°C). These temperatures were used to predict the $q$ on each interval. Root mean square error, bias error, and variance error are calculated in error analysis of the inverse technique. The mean squared error is used to check the accuracy of the estimation. This error [16] is calculated as

$$RMS = \sqrt{\frac{1}{N}\sum_{i=1}^{N}(\hat{q}_{i,noisydata} - q_i)^2} \qquad (3)$$

Where $\hat{q}_{i,noisydata}$ is the predicted heat flux. Bias error [16] is because of the regularization method and can be calculated as follows.

$$D = \sqrt{\frac{1}{N}\sum_{i=1}^{N}(\hat{q}_{i,nonoise} - q_i)^2} \qquad (4)$$

In which $\hat{q}_{i,nonoise}$ is estimated using temperatures with $\sigma = 0$. The inverse method's sensitivity to the measurement errors in the data is called the variance error [16] and can be calculated as:

$$V = RMS^2 - D^2 \tag{5}$$

For better comparison, the errors have the same scale, hence the second root of the variance is used here. It is necessary for every experimental test that an experimental design should be carried out. This work is also done in this study. In the inverse problem, it is quite clear that the sensors should be closer to the active surface (i.e. boiling surface) and the accuracy of this approach is higher due to the high sensitivity coefficient. In this experiment, the temperature sensors were located at $y = E - 0.5mm$ due to the precision of the machining equipment. In IHCP [21-22], a small time step causes an increase in the variance error and a very large time step increases the bias error. We are looking for a suitable time step that the RMS error is low and we can reach more details about the unknown parameter over time. The time step is determined according to the RMS. The time step is considered 0.01s, 0.05s, 0.1s and 0.5s and using the numerical simulation experiments, the error is calculated and is equal to values 0.8$q_{mean}$, 0.05$q_{mean}$, 0.069$q_{mean}$ and 0.072$q_{mean}$, respectively. According to the results, 0.1 seconds is selected. It should be noted that the smallness of the sensitivity coefficient has a direct effect on the increase of RMS error.

Table 4 presents the error analysis for estimated heat flux on the heat exchange surface. Results show that the proposed method has a low error, estimating heat flux. This error is almost 0.06 $q_{mean}$. In the proposed method, firstly the $q(x,t)$ is predicted and then, using Newton's cooling law, $h$ is calculated; therefore, the $h$ is estimated indirectly. The proposed algorithm is simple and easy to use, with its results showing it has a good accuracy.

**Table 4** Error analysis for the purposed inverse method

| Error | RMS(W/m²) | Bias(W/m²) | variance(W/m²) |
|---|---|---|---|
| Estimated heat flux | 0.069× q$_{mean}$ | 2.4e-14× q$_{mean}$ | 0.069× q$_{mean}$ |

**Results**

The $h(x,t)$ is calculated using the $q(x,t)$, itself estimated by the inverse method $(q(x,t))$ and the calculated local surface temperature $(T_s(x,t))$ as follows as:

$$h(x,t) = \frac{q(x,t)}{T_s(x,t) - T_f(x,t)} \tag{6}$$

Where $T_f(x,t)$ is the bulk temperature and is calculated, using the energy balance equation:

$$T_f(x,t) = T_f(x - \Delta x, t) + \frac{q(x,t)(2H + 2W)\Delta x}{\dot{m}C_P} \tag{7}$$

In which q and T$_s$ are estimated through solving IHCP, $H$ is channel height, and $W$ is channel width. The heat transfer coefficient ($Nusselt\ number$) [2] is specified as follow as:

$$Nu(x,t) = \frac{h(x,t)}{k}(2\frac{HW}{H+W}) \tag{8}$$

This study investigates $h(x,t)$ for flow with and without phase change. Several experiments were tested in this study. Considering the number of thermocouples and time-steps, millions of temperatures were acquired as raw data, which we couldn't present here. Figure 4 illustrates the measured temperatures at each position of the sensor in the flow without phase change (Re=1800) and in liquid-vapor phase flow $G = 355$ kg/(m²s). An increase in temperature is observed with increasing $x$. Also, temperature increases over time. As the flow of fluid enters the channel (at $x = 0$), the thermal boundary layer forms and develops. The Nusselt number in the channel entrance (at $x = 0$), is very large. With the growth of the thermal boundary layer along the channel, the Nusselt number reduces until the flow (thermal and hydrodynamic) is fully developed. At this time, the heat transfer coefficient does not change with $x$ and is constant. During the two phase flow, the temperature increases along the channel. First, the temperature increases with the time it starts to boil and the bubble is generated inside the stream, in this case, the temperature reaches a constant value. In fact, Measurements emphasize on this issue the essence of the flow inside the channel strongly affects surface temperature variations.

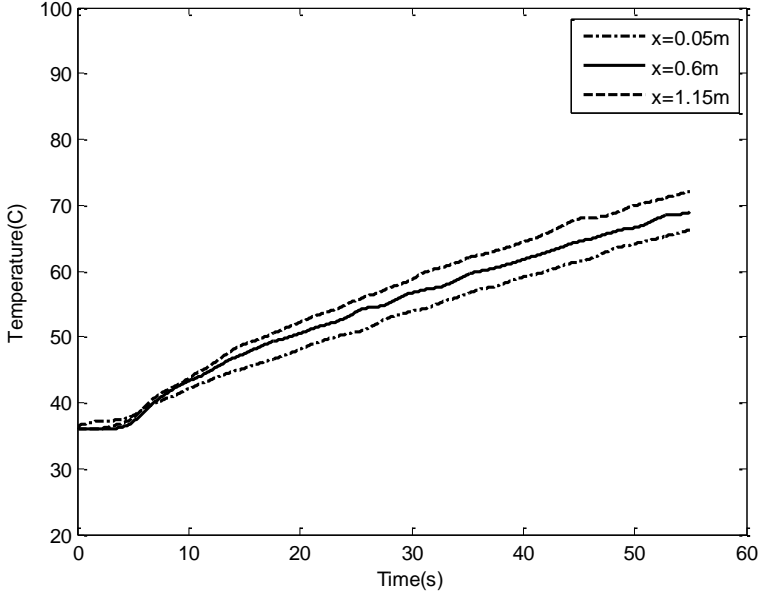

a)

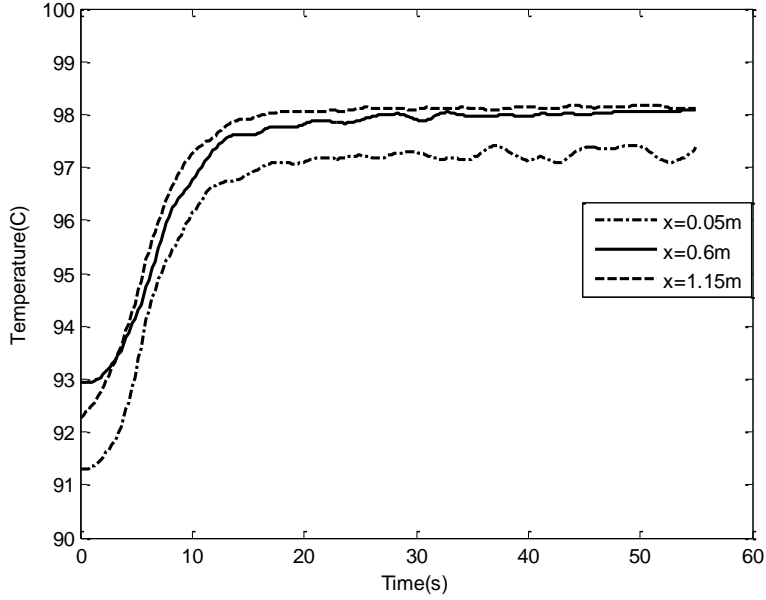

b)

**Figure 4** the measured temperatures a)single phase (Re=1800)and b)two-phase flow(G=355 kg/(m².s))

**Single-Phase Flow**

The effect of $Re$ changes on $Nu(x.t)$ in the single-phase flow has been studied. The obtained results for single-phase flow are validated with Equation (10). Churchill and Ozoe [29] used experimental results to find a correlation to predict the Nusselt number in a fully developed plug flow in a pipe with $q = cte$. A closed-form expression [29] for the $Nu_x$, covering both the developing and developed flow in a tube with $q = cte$ is:

$$\frac{Nu_x}{4.364[1+(Gz/29.6)^2]^{1/6}} = [1+(\frac{Gz/19.04}{[1+(Pr/0.0207)^{2/3}]^{1/2}[1+(Gz/29.6)^2]^{1/3}})^{3/2}]^{1/3} \quad (9)$$

$$Gz = \pi/(\frac{4x}{Re\,Pr})$$

For all Gz, this equation agrees within 5% error in developing flow and 3.5% error developed flow for Pr=0.7 and Pr=10[29]. In calculating the Nusselt number for non-circular cross sections, a hydraulic diameter is utilized. In these cross sections, the Wall-Stream effect must be considered in calculating the Nusselt number based on the hydraulic diameter $\left(D_h = \frac{2HW}{H+W}\right)$. This correction [30] is done using the coefficient of $\beta = ((\pi D_h^2)/4)/A_{duct}$. Based on the modification of obtained $Nu_x$ extracted from Eq. (9), it could be rephrased as [30]:

$$Nu_x = Nu_x \times \beta \tag{10}$$

The importance of β can be seen in Equation (10). Also, Figure (5) illustrates the estimated time distribution of heat transfer coefficient for Re= 1165 and Table 5 gives difference of estimated $Nu$ values with the values calculated from Equation (10), showing the highest and lowest difference to be 18% and 0.4%, respectively. A comparison of the estimated single-phase Nu with one predicted (eq.10) for laminar flow in rectangular channels is shown in figure 6a, wherein time-averaged estimated Nu is compared with the values, obtained from Equation (10).

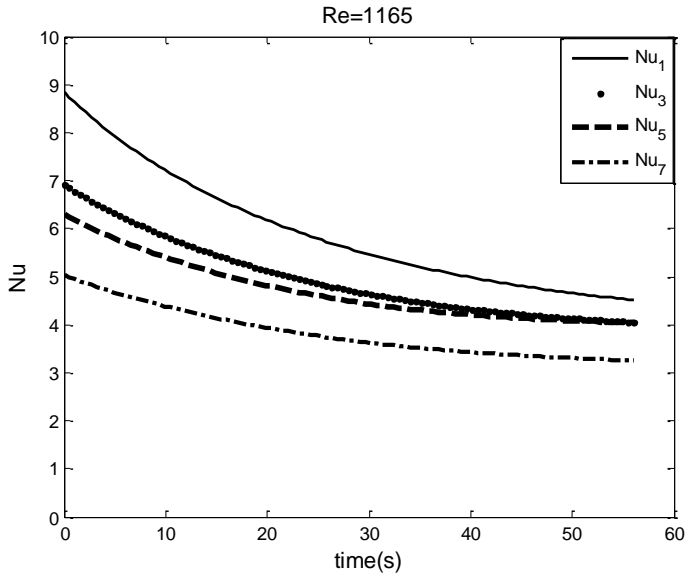

a)

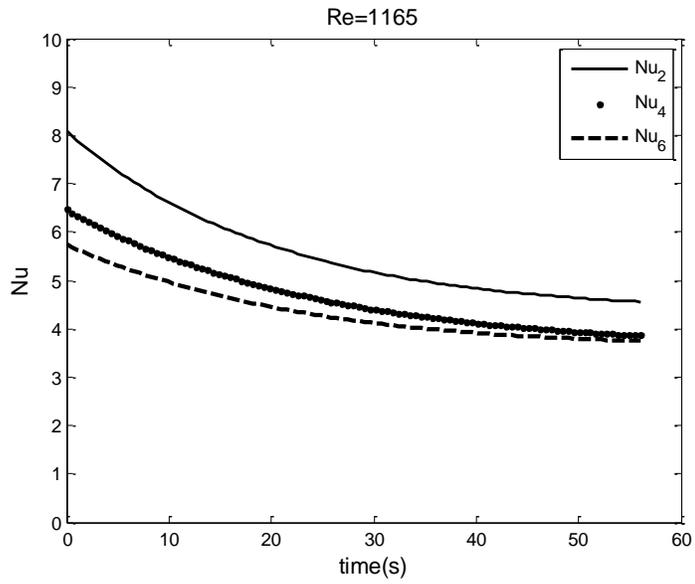

b)

**Figure 5** The estimated time history of Nusselt number of single-phase flow for Re=1165

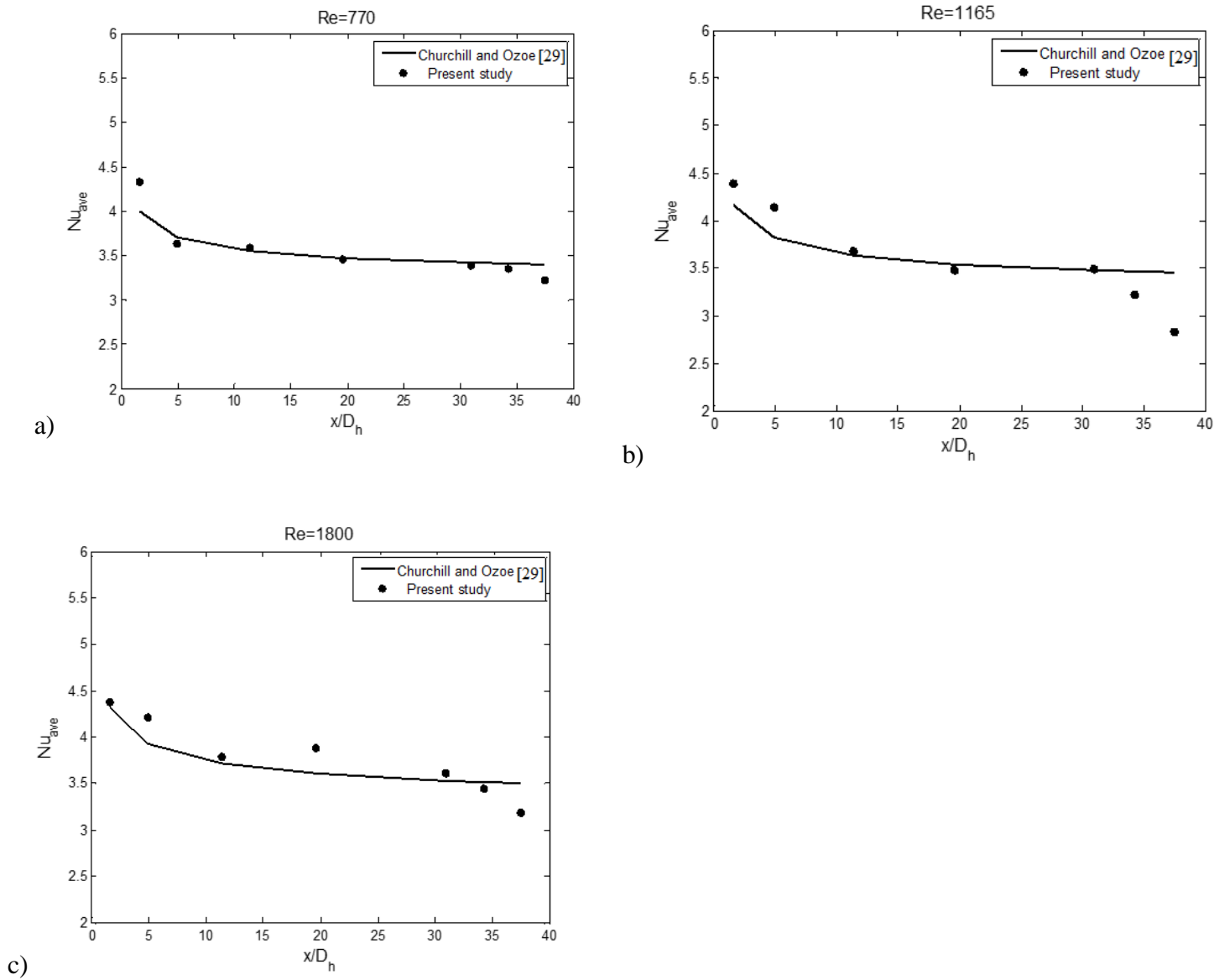

**Figure 6** The local Nusselt number along the channel length for a)Re=770, b)Re=1165 and c)Re=1800

At the present work, and the difference of values calculated from the two correlations is about 7 %, and the results are approximately one. Perhaps because the geometry is simple and this affects the flow structure. The $Nu_x$ for Re=770, 1165 and 1800 is illustrated in Figure6 . Furthermore, the averaged deviation percentage of estimated Nusselt number with the results of Equation (10) for Re=770, 1165 and 1800 is given in Table 6. The relative deviation is about 2%. Quite obviously, the reason for the deviations is the method for the prediction of the *Nusselt number*. Table 6 shows where the predicted values fully agree with equation (10). The mentioned method has a good accuracy to estimate heat transfer coefficient. It shows that the proposed method can successfully estimate Nusselt number for flow without phase change.

**Table 5 Deviation** between the extracted Nu from eq.10 and estimated Nu

| Location(m) | Nu | Nu in present study | deviation |
| --- | --- | --- | --- |
| 0.050 | 4.450 | 4.220 | -0.054 |
| 0.150 | 4.197 | 3.870 | -0.084 |
| 0.350 | 3.729 | 3.680 | -0.013 |
| 0.600 | 3.523 | 3.588 | 0.018 |
| 0.950 | 3.537 | 3.523 | -0.004 |
| 1.050 | 3.270 | 3.510 | 0.068 |
| 1.150 | 2.873 | 3.499 | 0.178 |

**Boiling Flow**

Table 7 gives the experimental conditions for boiling flow. The mass flux varies between 118.1 and 355 $kg/(m^2.s)$, whereas the $q_w$ is from 0 to 19.2 $kW/m^2$. Figure 7 shows time variations of the estimated $q(x,t)$ on the heat exchange surface for G=234$kg/(m^2 s)$. Nusselt number for boiling flow is calculated using equation (6), in which $T_f(x.t)$ will be equal to saturation temperature, $T_{sat}$.

**Table 6** Mean deviation between the estimated Nu and calculated Nu from Equation(10)

|  | Re=1165 | Re=1800 | Re=770 |
|---|---|---|---|
| **Equation(10)** | 3.60 | 3.78 | 3.56 |
| **estimated** | 3.64 | 3.73 | 3.56 |
| **Error** | 0.011 | -0.014 | -0.001 |

**Table 7** Mass flux and heat flux for two phase flow

| G( kg/(m².s)) | 118.1 | 234 | 355 |
|---|---|---|---|
| $q_w$(kW/m²) | 7.9 | 14 | 18.9 |

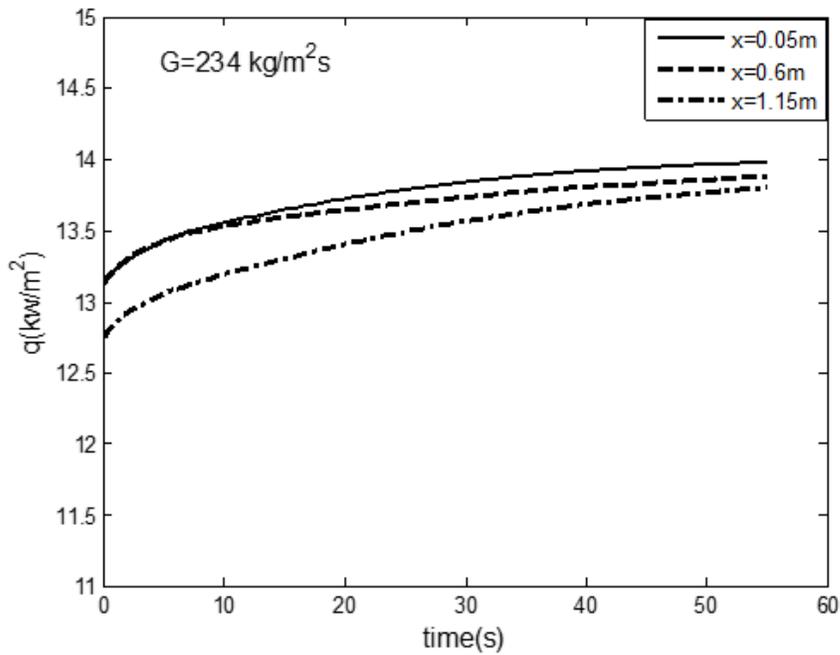

a)

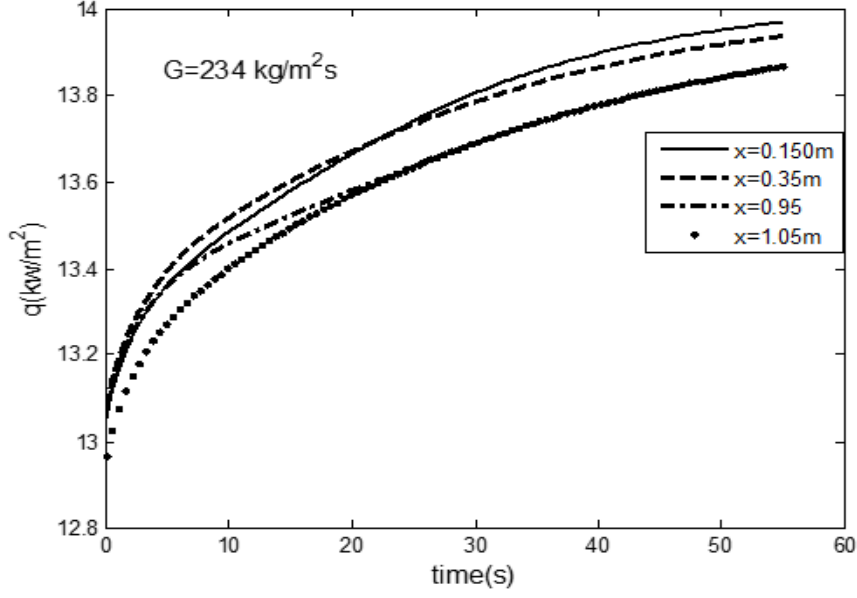

b)

**Figure 7** The time history of the local heat flux along the channel length

To compare the Nusselt number of boiling flow inside the tube, it has been applied to the correlation that Kandlikar [31] has been given. Kandlikar's correlation [31] is

$$Nu = Nu_{sp}[0.6683Co^{-0.2}(25\frac{G^2}{\rho_L^2 gD_h})^{0.3} + 1058(\frac{q}{Gh_{fg}})^{0.7}] \tag{11}$$

Where $Nu_{sp}$ is Nusselt number in the single-phase state and is calculated by assuming that all flow inside the channel is liquid and Co is the convection number, defined by [31]:

$$Co = (\frac{1-x_v}{x_v})^{0.8}(\frac{\rho_G}{\rho_L})^{0.5} \tag{12}$$

In which $x_v, \rho_L,$ and $\rho_G$ are the local vapor quality, the density of liquid and vapor, respectively. On average, the deviation of this equation for water and all the refrigerants is about 15.9% and 18.8%, respectively [31]. In order to specify the quality of vapor, the energy equilibrium equation for every section between the input and the output is employed. The vapor mass qualities [31] is calculated as:

$$X_v(x,t) = X_v(x-\Delta x,t) + \frac{1}{h_{fg}}[2\frac{q(x,t)(H+W)}{\dot{m}}\Delta x - C_P(T_{sat} - T_f(x,t))] \quad (13)$$

Figure 8a shows the time-averaged Nu decreases with increasing $x$ and the experimental results are close to the predictions of Kandlikar [31]. Near the entrance of channel is the difference between the estimated values and the reference values [31] at about 19 % and in the rest of the regions, this difference is insignificant. The reason for this difference between the empirical correlations with the present work can be attributed to how the heat flux is calculated at the $y = E$, in these two empirical correlations the authors have assumed that $q(x.t) = q_w$. Another reason may be that the curve fitting of the experimental data has errors and the results of the correlation on the test data are not fully matched.

Time distribution of vapor quality along the channel is illustrated in Figure 8 b. The quality of the vapor increases in the flow direction along the channel, thanks to the bubble coalescence mechanism; therefore, the frequency of the bubbles is reduced. This explains why heat transfer is reduced when the quality of steam is increased. It can be determined when the vapor generation along channel length starts, as it is one of the findings of the proposed method. The heat transfer of boiling flow consists of the convective and the nucleate boiling terms. Using the proposed method, the role of each component can be determined.

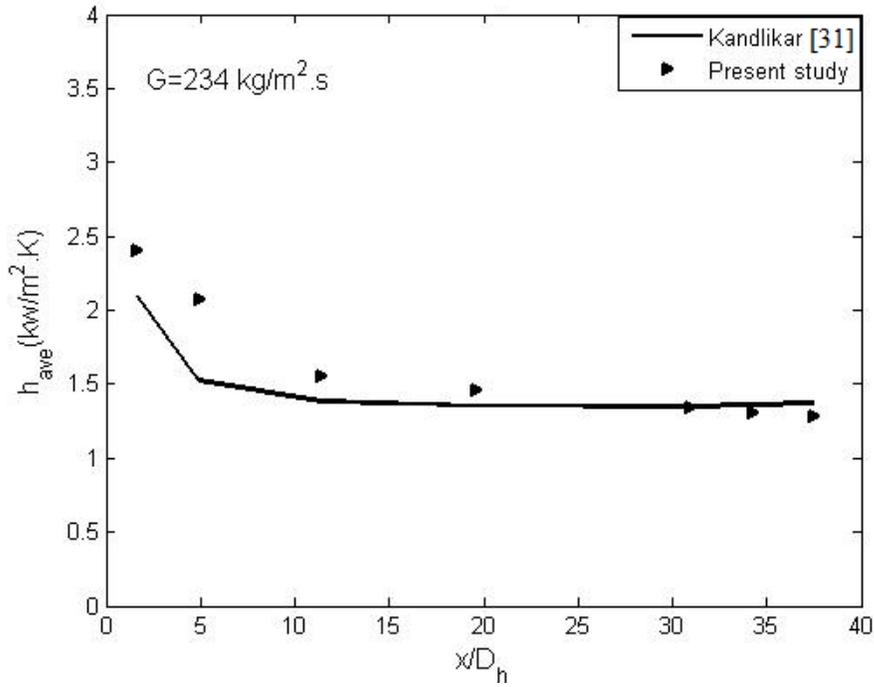

a)

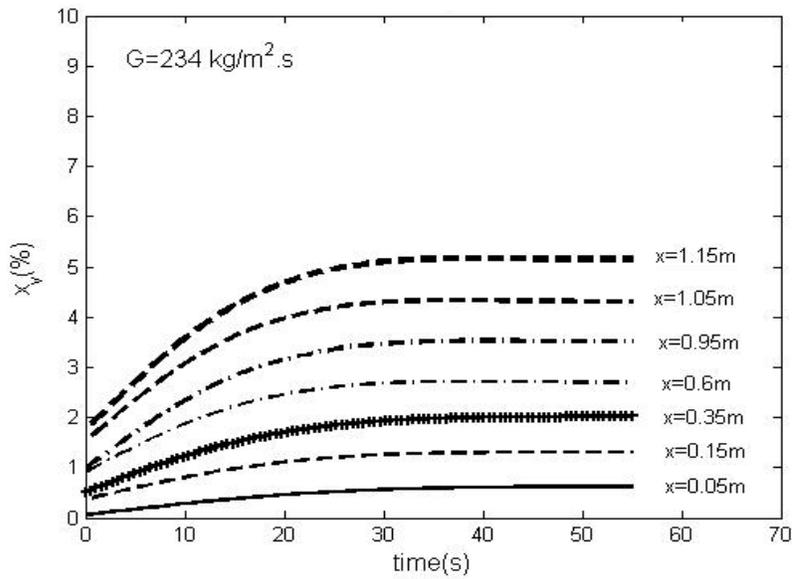

b)

**Figure 8** a) the local convective heat transfer coefficient along the channel length, and b) time variation of local vapor quality along the channel length

Figures (9) and (10) demonstrate the variations of *Co number* and *Bo number* with time. Convection number decreases with time, similar to the trend of changes in Convection number along the channel, which descends as well. The maximum amount of Co corresponds to the 50 mm distance from the channel entrance. Boiling number at the measurement points increases with time. Temporal and spatial Variations of Boiling number is similar to the variations of the predicted heat flux. The nucleate boiling heat transfer is dominant when the $Co$ number is too large. In the nucleate boiling region, the $Nu$ is a weak function of the convection number because the mechanism of convection has little effect in this area and is not the dominant mechanism. In the area where the convective boiling mechanism is dominant, the amount of convection number is small. The contribution of the convective boiling decreases when the Boiling number is increased. Figure 11 illustrates the variations in the time-averaged of Nusselt number for G = 118.1,234, and 355 kg/(m².s) with the quality of vapor. As the figure shows the Nusselt number reduces with an enhancement in $X_v$. The $X_v$ is low when the nucleate boiling is dominant. This mechanism and convective heat transfer help to enhance the heat transfer. The time distribution of local Nusselt number for G=118.1 kg/(m².s) is presented in Figure 12. At x = 50 mm the number varies with time but does not change much in other measurement points.

a) 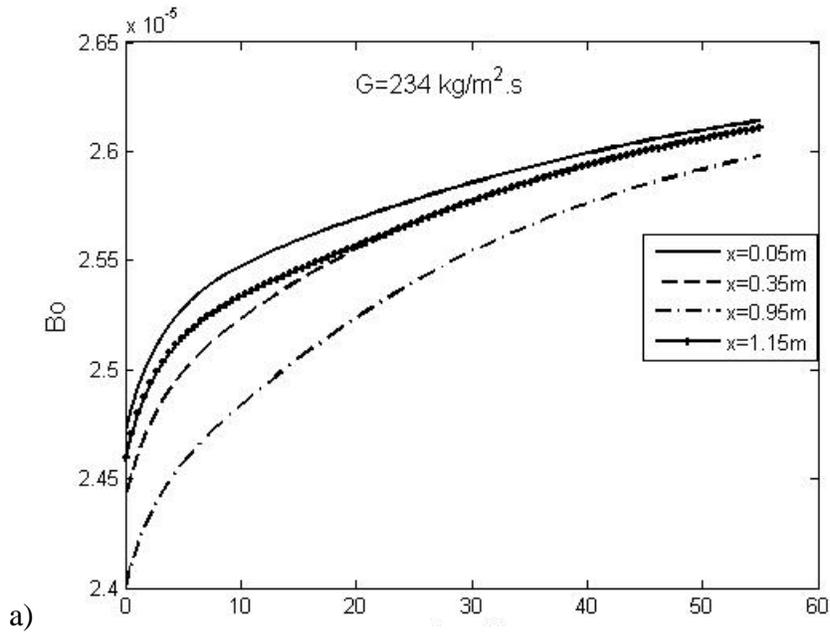

b) 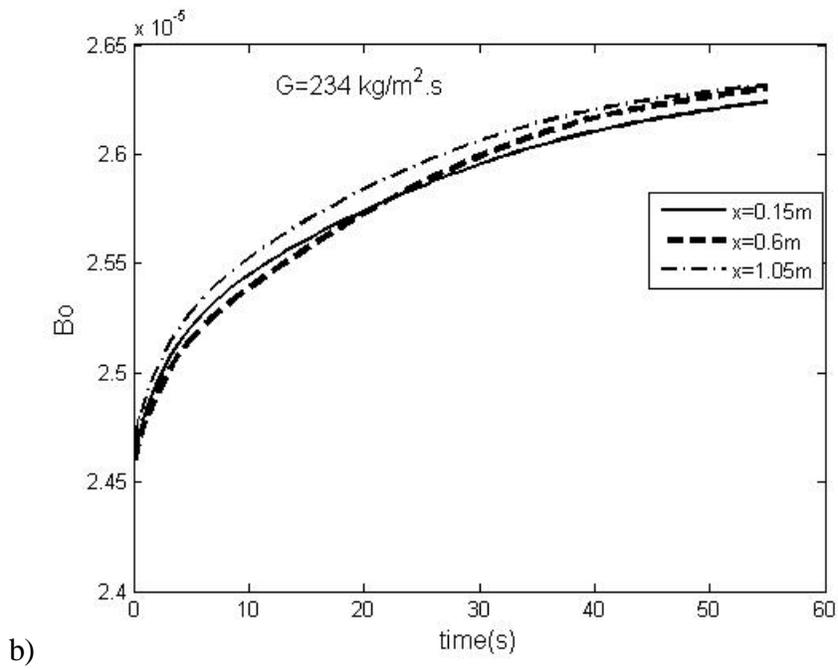

**Figure 9** The variations of Boiling number (Bo) with time along the channel

a)

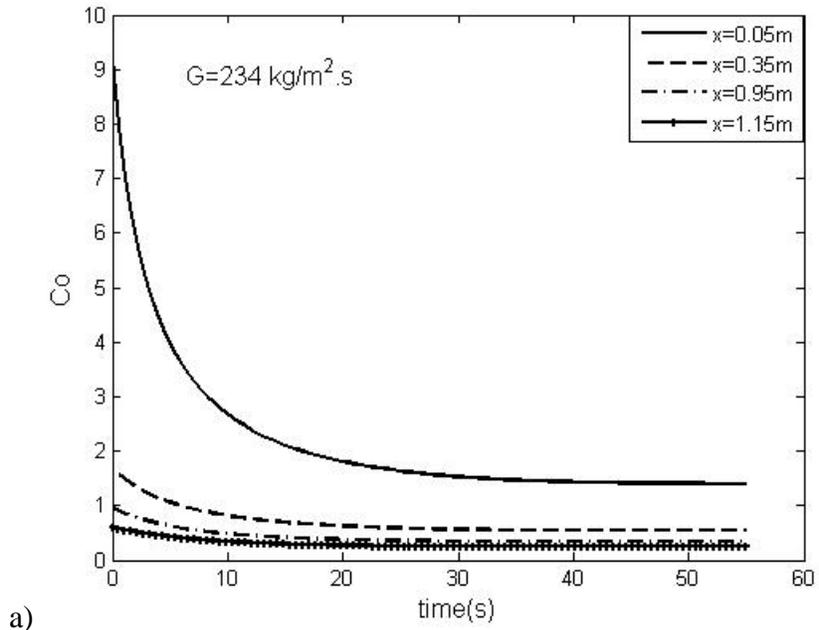

b)

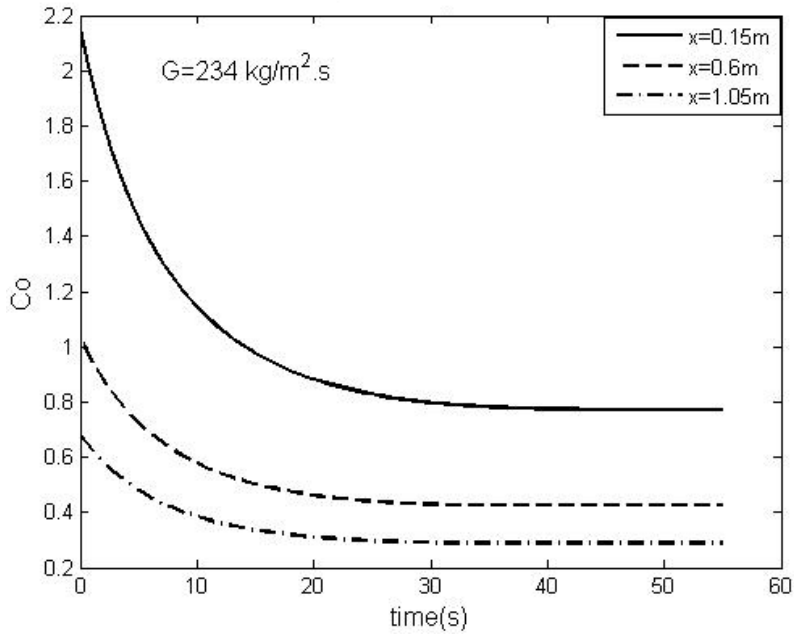

**Figure 10** The variations of convection number (Co) with time along the channel

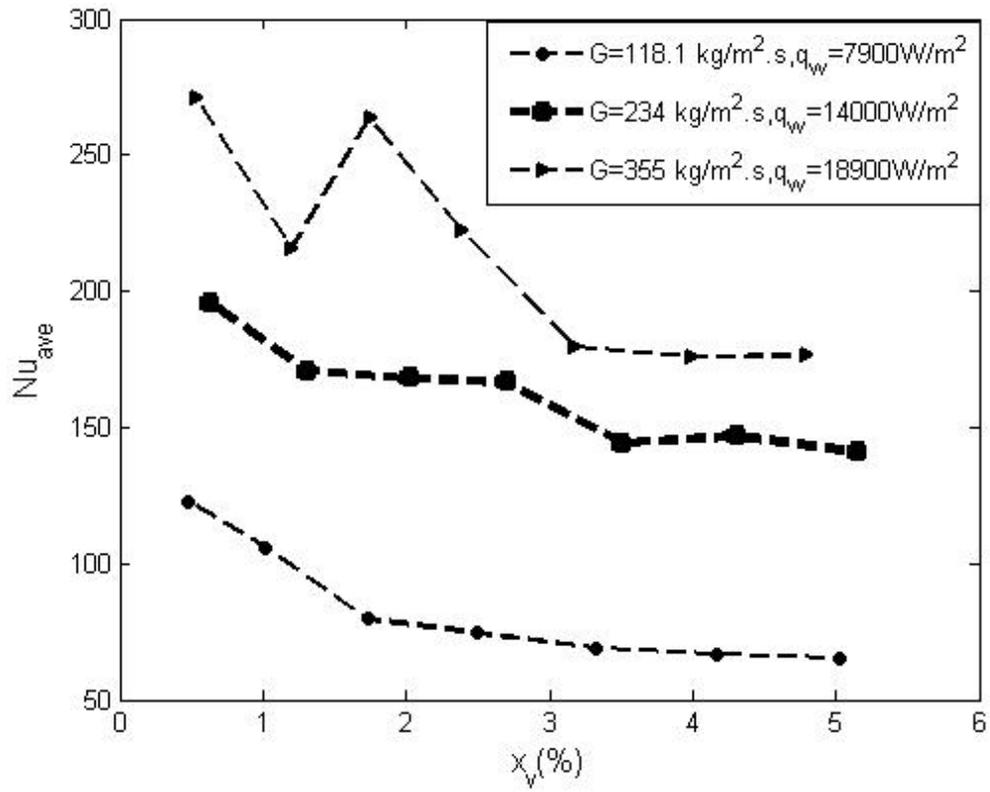

**Figure 11** Time-averaged heat transfer coefficient with the vapor quality

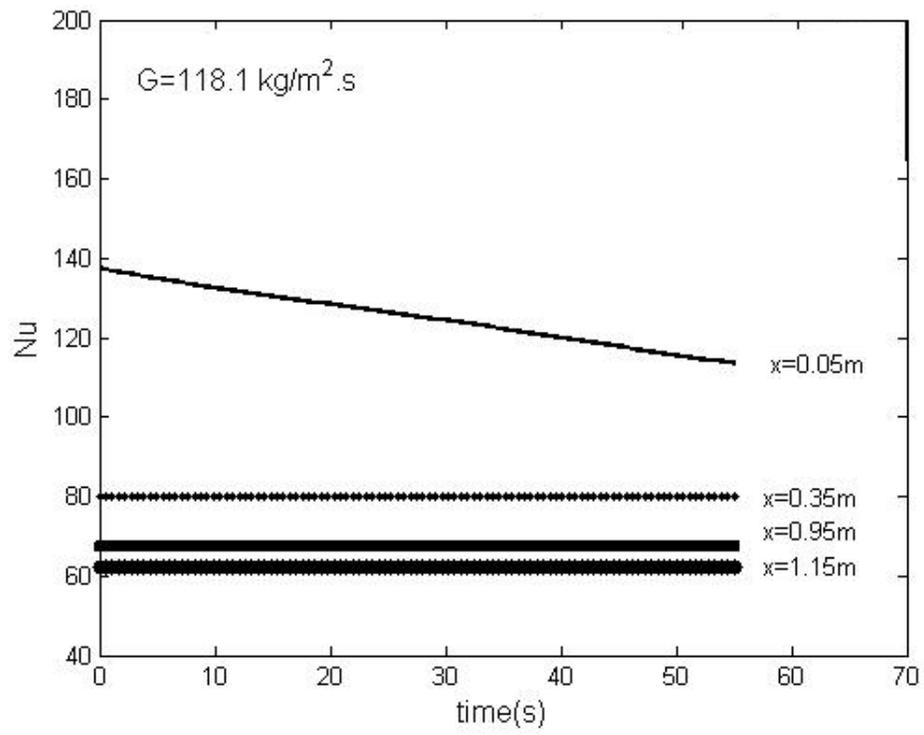

a)

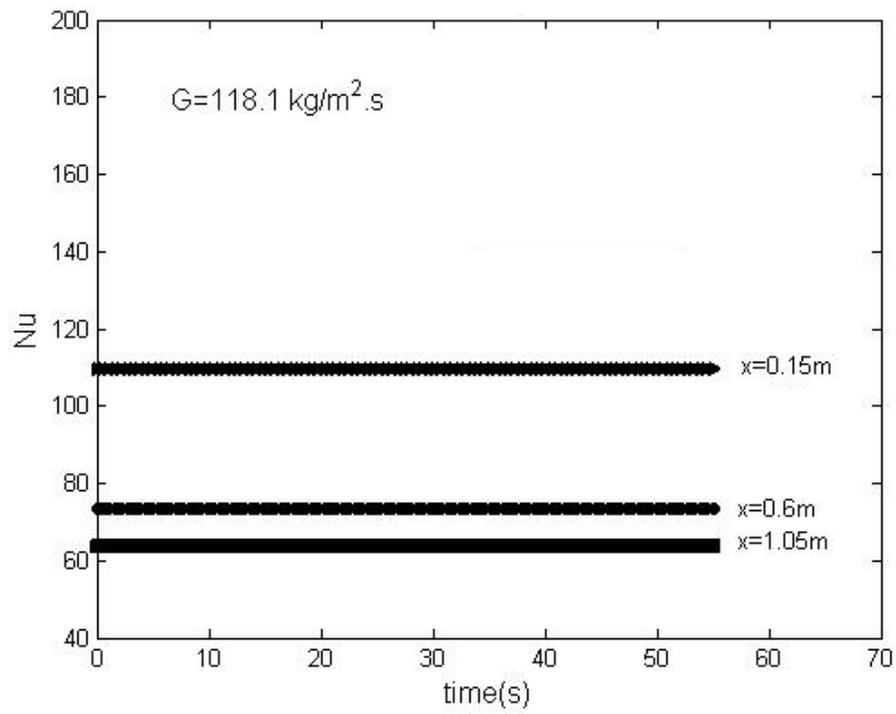

b)

**Figure 12** The time variation of local Nusselt number for G=118.1 kg/(m$^2$.s)

**Conclusions**

This research conducted a laboratory investigation of water flow with and without phase change in a channel. The temperatures within the wall were experimentally measured, providing the input data for inverse method to predict the $q(x,t)$. Transient analysis was performed. First, the experiment was designed and error analysis carried out to determine the accuracy of the proposed method. RMS error for estimation heat flux was around 0.06q$_{mean}$. Using Newton's cooling law and estimated values of heat flux, temporal and spatial distribution of Nusselt number was predicted. The time-averaged local Nusselt number estimated for flow with and without phase change were close to those, predicted by the correlations of Churchill and Ozoe [29] and Kandlikar [31], respectively. It was seen in boiling flow that local quality of vapor and boiling number increased and local convection number decreased with time. Also, along the length of channel time-averaged Nusselt number decreased with quality of vapor. The temporal and spatial changes of the Convective number and the Boiling number can be determined using the proposed method. This helps better understanding the boiling phenomenon.

**Nomenclature:**

Bo    Boiling number, dimensionless

Co    Convection number, dimensionless

d     Conjugate direction used in figure(3)

D     bias error

G     Mass flux, kg/(m².s)

Gz    Graetz number, dimensionless

H     height of channel, m

h     convective heat transfer coefficient, W/(m²K)

K     thermal conductivity, W/(mK)

E     length of plate, m

L     thickness of plate, m

N     measurements number

Np    unknown parameter numbers

Ns    sensor numbers

Nu    Nusselt number, dimensionless

$Nu_{sp}$  single-phase Nusselt number, dimensionless

$\dot{m}$   mass flow rate, kg/s

Pr     Prandtl number, dimensionless

q     heat flux, W/m$^2$

$q_w$     Heater's heat flux, W/m$^2$

Re     Reynolds number, dimensionless

RMS     root mean square error

T     calculated temperatures, K

$T_{sat}$     saturation temperature, K

$T_f$,     local fluid temperature, K

V     variance error

$X_v$     quality of vapor

Y     measurements, K

W     width of channel, m

Greek Symbols

$\rho$     density, kg.m$^{-3}$

$\alpha$     thermal diffusivity, m$^2$/s

$\varepsilon$     value of noise

$\beta$     search step (figure (3))

$\gamma$     The conjugate coefficient (figure (3))